\def\aj{{AJ}}

\def\apj{{ApJ}}

\def\mnras{{MNRAS}}

\documentclass[cup5b]{caps}
\usepackage{graphicx}
\usepackage{amssymb}
\usepackage{ociwsymp3e}  

\HeadText{B. Ziegler et al.} 

\newcommand{\cl}{CL\,0413--65}
\newcommand{\ms}{MS\,1008--12}

\begin{document}

\pagenumbering{arabic}


%
%

\author[]{B. L. ZIEGLER$^{1}$, A. B\"OHM$^{1}$, K. J\"AGER$^{1}$,
A. FRITZ$^{1}$, and J. HEIDT$^{2}$
\\
(1) Universit\"ats--Sternwarte, G\"ottingen, Germany\\
(2) Landessternwarte, Heidelberg, Germany\\}

\chapter{Internal Kinematics of Spiral \\  Galaxies in Distant Clusters}

\begin{abstract}
We present first results from our project to examine the internal
kinematics of disk galaxies in 7 rich clusters with $0.3\le z<0.6$. Spatially
resolved MOS spectra have been obtained with FORS at the VLT. We concentrate 
here on the clusters MS\,1008.1--1224 at $z=0.30$ and 
Cl\,0413--6559 (F1557.19TC) at $z=0.51$. Out of 22 cluster
members, 12 galaxies exhibit a rotation curve of the universal form 
rising in the inner region and passing over into a flat part. The other 
members have intrinsically peculiar kinematics.
The 12 cluster galaxies for which a maximum rotation velocity could be derived 
are distributed in the Tully--Fisher diagram very similar to field 
galaxies from the FORS Deep Field with corresponding redshifts. The same is
true for 6 galaxies observed in the cluster fields that turned out not 
to be members.
In particular, these cluster spirals do not show any significant luminosity 
evolution as might be expected from certain clusterspecific phenomena.
Contrary to that, the other half of the cluster sample with disturbed 
kinematics also shows a higher degree of structural assymetries on average
indicating ongoing or recent interaction processes.
\end{abstract}

\section{Introduction}

Galaxies in clusters experience various interactions, some mechanisms of which
can cause substantial distortions of the internal kinematics of disk 
galaxies leading to ``rotation curves'' that no longer follow the universal 
form (Persic et al. 1996)
of spirals in the field. In a study of 89 galaxies in the Virgo cluster,
Rubin et al. (1999), for example, classified half of their sample galaxies
as kinematically disturbed ranging from modest (e.g. asymmetric) to severe
(e.g. truncated curves) peculiarities. On the other hand, many cluster 
galaxies, for which only velocity widths (in contrast to spatially resolved
velocities) were measured, follow a tight Tully--Fisher relation (e.g. 
Giovanelli et al. 1997). Furthermore, it is not yet clear whether the halo of 
dark matter and, therefore, the total mass of a galaxy can also be affected by 
certain interaction phenomena. In numerical simulations of cluster evolution
that can resolve substructure and use semianalytic recipes to describe the
galaxies' stellar populations, the dark matter halo of a galaxy that falls into
the cluster does survive until redshift
zero and does not get stripped (Springel et al. 2001).

A significant contribution to our comprehension of the environmental dependence
of galaxy kinematics would be made by observations at higher redshifts, since 
in models of hierarchically growing structure clusters are still in the 
process of forming at $z\approx1$ so that a higher infall rate and more 
interactions are expected (e.g. Kodama \& Bower 2001). 
A first (published) attempt was made by Metevier et al. (2002 \& this
conference) with 10 Tully--Fisher galaxies in the cluster 
CL\,0024$+$1654 ($z=0.40$) and by Milvang--Jensen et al. (2003) with 8 spirals
in MS\,1054.4--0321 ($z=0.83$).
While the first study sees a larger scatter of their sample galaxies in the TF
diagram without evidence for an evolution of the zero point, the second one finds a trend towards
brighter luminosities (0.5--1 $B$ mag) with respect to local cluster spirals. 
It is argued that either truncation
of star formation or starbursts may cause an increased or decreased 
mass--to--light ratio, respectively.

\section{Project aims and observations}

We have conducted a campaign (5 nights with
FORS\,1\&2 at the ESO--VLT) to gain spectra of spiral galaxies within 
seven distant clusters in the redshift range $0.3\le z<0.6$. 
One of our main goals is the derivation of spatially resolved rotation curves 
for the analysis of the internal kinematics of the galaxies
and the construction of the distant cluster TFR.
Furthermore, we aim to determine the star formation rates, structural 
parameters and morphologies to investigate possible effects of transformation 
mechanisms like merging, tidal interaction, harassment, strangulation or 
interaction with the intra--cluster medium. 
To complement our own ground--based imaging, HST/WFPC2 images 
of the core regions were retrieved from the archive.

  \begin{figure}
  \hspace*{-1.0cm}
    \includegraphics[width=13cm,angle=0]{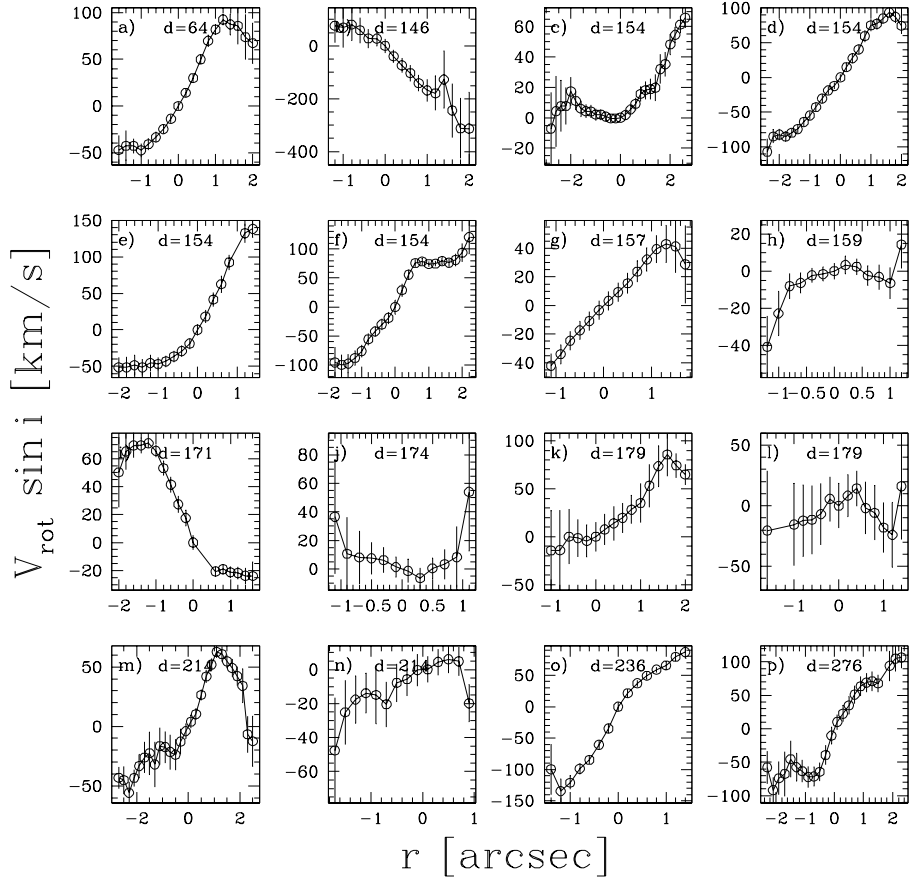}
     \caption{Position--velocity diagrams for 12 disk galaxies in the 
   cluster \ms\ at $z=0.30$. Note, panels c\&d, e\&f, k\&l and m\&n 
   represent the same galaxy, rsp., but observed with different slit positions
   (the rotation angles of the two MOS masks differed by 67$^\circ$).
   Panels a--p are ordered according to the distance $d$ (given in arcsec) of
   the galaxies from the cluster center.
   Values of the maximum rotation velocity could be determined robustly for 
   cases a, d, e, f, g, i, o \& p. 
   These galaxies enter the TFR shown in 
   Fig.\,1.2. In all other cases, the kinematics are too disturbed. }
    \label{MS1008_RCs}
  \end{figure}

Here, we present first results based on data of two of the clusters,
\cl\ ($z=0.51$) and \ms\ ($z=0.30$), which were observed in November 1999 and  
March 2000, respectively. 
We used the FORS1 instrument  
mounted at the Cassegrain focus of the ESO--VLT 
with grism 600R providing a spatial scale of 0.2 arcsec/pixel 
and a dispersion of $\sim$1.08 \AA/pixel.
At a slitwidth of 1\,arcsec, the spectral resolution was
$R\approx1200$. To meet our signal--to--noise requirements, the total
integration time was set to 2.5\,hrs. Seeing conditions ranged between 0.7 and 1.3
arcsec FWHM. Two MOS setups have been observed
for each cluster, yielding 76 spectra of galaxies with apparent magnitudes
of $18.0<R<22.8$.

\section{Derivation of rotation velocities} 

Redshifts could be determined for 62 of the galaxies,
of which 38 have late--type SEDs, 19 are early--type, and 5 peculiar or
unidentified. In total, 38 objects are cluster members.
22 of these are spirals with sufficient signal--to--noise of their
emission lines to derive the internal kinematics.

Rotation curves (RCs) were determined using either the  
[O\,{\sc ii}]3727, H$\beta$ or [O\,{\sc iii}]5007 emission line (Fig.\,1.1
gives some examples).
Gaussian fits have been applied stepping along the spatial axis
with a median filter window of typically 0.6 arcsecs to enhance the S/N. 
Since the apparent disk sizes of spirals at intermediate redshifts are
only slightly larger than the slit width (1\,arcsec), the spectroscopy
covers a substantial fraction of the two--dimensional
velocity field. Thus, the maximum rotation velocity $v_{\rm max}$ 
(the constant rotation in the outer part of a galaxy
due to the Dark Matter halo) cannot be determined ``straightforward'' from
the observed rotation.
To tackle this problem, we simulated the spectroscopy of each galaxies' 
velocity field with the respective inclination and position angle, also
taking seeing and luminosity weighting into account (see Ziegler et al. 2001
for a more detailed description). 
The simulated rotation curves which best reproduced the
observed ones yielded the values of $v_{\rm max}$ entering our TF diagram
(Fig.\,1.2).
  
  \begin{figure}
    \centering
    \includegraphics[width=10cm,angle=0]{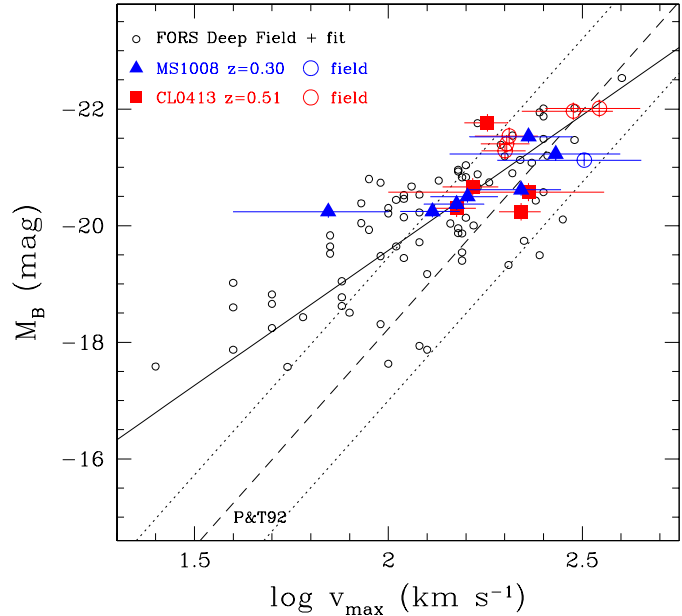}
    \caption{Tully--Fisher diagram of cluster spirals in \ms\ at $z=0.30$ 
   (triangles) and in \cl\ at $z=0.51$ (squares). 
   Also shown are six field
   objects (large open circles) that were also observed in the clusters' 
   field--of--view. In comparison to our FORS Deep Field sample of 77 field
   galaxies (Ziegler et al. 2001, B\"ohm et al. 2003)
   with a mean redshift of 0.5 (small open circles), the cluster
   galaxies are similarly distributed and do not deviate significantly from
   the linear fit to the FDF sample (solid line).
   The cluster members follow the same trend with respect to the local TFR
   (the fit $\pm3\sigma$ to the Pierce \& Tully 1992 sample is given)
   as the distant field galaxies. Since we observed mostly bright cluster
   galaxies, their luminosities are not significantly increased in
   accordance with an undisturbed evolution.
   (Restframe $B$ magnitudes were calculated for a flat $\Omega_\lambda=0.7$ 
   cosmology with $H_0=70$\,km\,s$^{-1}$\,Mpc$^{-1}$.)
   }
    \label{MS1008_RCs}
  \end{figure}

\section{Results and discussion}

For our sample, a significant number of the cluster members show peculiar 
kinematics in contrast to the ``rise--turnover--flat'' RC shape of 
large, isolated spirals. 
In the case of \ms\ (Fig.\,1.1), no trend is visible of the RC form with
(projected) clustercentric distance. But since all observed galaxies are 
located within the virial radius, this is in accordance with dynamical 
models in
which the galaxy population of a cluster is well-mixed within that region.
In particular, we most probably do not have any new arrivals from the field
in our sample. 

We point out that the peculiarities of some ``rotation cuves'' 
reflect mainly intrinsic kinematic properties 
of the  resp. galaxy.
This is in contrast to the significantly smaller fraction of 
peculiar curves in our sample of 77 distant FORS Deep Field spirals, 
which have been observed with exactly the same instrument configuration
(Ziegler et al. 2001, B\"ohm et al. 2003).
Only those rotation curves which show no strong perturbations are eligible
for a determination of the maximum rotation velocity as needed
for the TF diagram. 
In the case of \ms, this could be done for 7 cluster members,
in \cl\ for 5 members. Thus, in total 12 cluster galaxies enter our
TF diagram which is shown in Fig.\,1.2. 
The luminosities  were derived from total magnitudes of
FORS images in the $V$ (\ms) and $I$ (\cl) band, rsp., corrected for Galactic
and intrinsic extinction, transformed to restframe Johnson $B$ according to 
SED type, and calculated for a flat $\Omega_\lambda=0.7$ cosmology 
($H_0=70$\,km\,s$^{-1}$\,Mpc$^{-1}$).
 
An inspection of our preliminary TF diagram reveals that the distant cluster 
spirals are distributed very much alike the field population that covers
similar cosmic epochs ($\langle z_{\rm FDF}\rangle=0.5$). 
No significant deviation from the distant field TFR is visible and the cluster
sample has not an increased scatter, but the low number of cluster members 
prohibits any quantitative statistical analysis. Nevertheless, we can conclude
that
the mass--to--light ratios of the observed distant cluster spirals cover the
same range as the distant field population indicating that no clusterspecific
phenomenon dramatically changes the stellar populations. In particular, there
was no starburst in the recent past of the examined cluster galaxies that would
have significantly risen their luminosity.
With respect to the Tully--Fisher relation obeyed by local galaxies
(e.g. Pierce \& Tully 1992), our cluster sample follows the same trend as the
FDF galaxies. Since we mostly selected bright galaxies, the cluster members
occupy a region in the TF diagram where no significant luminosity evolution
is visible.

But we emphasize that this conclusion is true only for those objects that
enter the TF diagram. Since almost half of our cluster galaxies can not be
used for a TF analysis due to their disturbed kinematics the above conclusions
are not generally valid for the whole cluster sample.
The objects with peculiar velocity curves may actually be subject to ongoing or
recent interactions. Indeed, a morphological classification of the cluster 
spirals in terms of their concentration and asymmetry indexes (Abraham et al. 
1996) reveals that they have a higher degree of asymmetry in the mean than 
the kinematically less disturbed ones. 
The asymmetric spirals show peculiarities like diverging rotation curves
or no significant rotation at all.
In one case (panels m\&n) of Fig.\,1.1), we observe a prominent bar that 
causes a highly asymmetric rotation curve.
This correlation between morphological and kinematical disturbance
hints to a common origin for both
in ``strong'' interactions like close encounters with
tidal effects, accretion events or even mergers. Such processes most probably
also influence the stellar populations of a galaxy changing its
integrated luminosity as well.

A more detailed discussion on the efficiency of the various interaction 
mechanisms,
however, needs to be put on the basis of robust statistics and will be
given for our complete sample of 7 clusters in future papers.

\vspace*{0.5cm}

{\bf{Acknowledgement:}} Based on observations collected with the VLT--telescopes
on Cerro Paranal (Chile) operated by the European Southern Observatory
(64.O--0158 \& 64.O--0152).
This work has been supported by the Volkswagen Foundation (I/76\,520)
and the Deutsche Forschungsgemeinschaft (Fr 325/46--1 and SFB 439).



\begin{thereferences}{}

\bibitem{AB}
Abraham et al. 1996, \mnras, 279, L47


\bibitem{Bo}
B\"ohm, A., Ziegler, B. L., \& Fricke, K. J., et al. 
2003, in The Evolution of Galaxies
III. From Simple Approaches to Self-Consistent Models, ed. G. Hensler 
(Dordrecht: Kluwer), in press (astro--ph/0210317)

\bibitem{GI}
Giovanelli, R., Haynes, M. P., Herter, T., Vogt, N. P., Wegner, G., 
Salzer, J. J., da Costa, L. N., \& Freudling, W., 1997, \aj, 113, 22


\bibitem{KO}
Kodama, T., \& Bower, R. G. 2001, \mnras, 321, 18

\bibitem{ME}
Metevier, A., Koo, D., \& Simard, L. 2002, in Tracing Cosmic Evolution with 
Galaxy Clusters, eds. S. Borgani, M. Mezzetti, \& R. Valdarnini (San Francisco:
ASP), 169

\bibitem{MI}
Milvang-Jensen, B., Arag\'on-Salamanca, A., Hau, G. K. T., J{\o}rgensen, I.,
\& Hjorth, J. 2003, \mnras, 339, L1

\bibitem{PS}
Persic, M., Salucci, P., \& Stel, F. 1996, \mnras, 281, 27

\bibitem{PT}
Pierce, M. J. \& Tully, R. B. 1992, \apj, 387, 47

\bibitem{RU}
Rubin, V. C., Waterman, A. H., Kenney, J. D. P. 1999, \aj, 118, 236

\bibitem{SP}
Springel, V., White, S. D. M., Tormen, G., \& Kauffmann, G. 2001, \mnras, 328,
726

\bibitem{BZ}
Ziegler, B. L., B\"ohm, A., Fricke, K. J., J\"ager, K. J., et al. 2002, \apj, 
564, L69


\end{thereferences}
\end{document}